\documentclass[nojss]{jss}
\usepackage{booktabs}
\usepackage[toc,page]{appendix}

\newcommand{\CRANpkg}[1]{\pkg{#1}}
\RequirePackage{fancyvrb}
\RequirePackage{alltt}
\DefineVerbatimEnvironment{example}{Verbatim}{}
\shortcites{sciencecloud,janus,dremel,nlme}
\author{Dirk Eddelbuettel\\Debian Project \And 
        Murray Stokely\\Google, Inc \And
        Jeroen Ooms\\UCLA}
\title{\pkg{RProtoBuf}: Efficient Cross-Language Data Serialization in \proglang{R}}

\Plainauthor{Dirk Eddelbuettel, Murray Stokely, Jeroen Ooms} 
\Plaintitle{RProtoBuf: Efficient Cross-Language Data Serialization in R}
\Shorttitle{\CRANpkg{RProtoBuf}: Protocol Buffers in \proglang{R}} 

\Abstract{
Modern data collection and analysis pipelines often involve
a sophisticated mix of applications written in general purpose and
specialized programming languages.  
Many formats commonly used to import and export data between
different programs or systems, such as \texttt{CSV} or \texttt{JSON}, are
verbose, inefficient, not type-safe, or tied to a specific programming language.
Protocol Buffers are a popular
method of serializing structured data between applications---while remaining
independent of programming languages or operating systems.
They offer a unique combination of features, performance, and maturity that seems
particulary well suited for data-driven applications and numerical
computing.
The
\CRANpkg{RProtoBuf} package provides a complete interface to Protocol
Buffers from the
\proglang{R} environment for statistical computing.
This paper outlines the general class of data serialization
requirements for statistical computing, describes the implementation
of the \CRANpkg{RProtoBuf} package, and illustrates its use with
example applications in large-scale data collection pipelines and web
services.
}
\Plainkeywords{R, Rcpp, Protocol Buffers, serialization, cross-platform} 

\Address{
  Dirk Eddelbuettel \\
  Debian Project \\
  River Forest, IL, USA\\
  E-mail: \email{edd@debian.org}\\
  URL: \url{http://dirk.eddelbuettel.com}\\
  \\
  Murray Stokely\\
  Google, Inc.\\
  1600 Amphitheatre Parkway\\
  Mountain View, CA, USA\\
  E-mail: \email{mstokely@google.com}\\
  URL: \url{http://www.stokely.org/}\\
  \\
  Jeroen Ooms\\
  UCLA Department of Statistics\\
  University of California\\
  Los Angeles, CA, USA\\
  E-mail: \email{jeroen.ooms@stat.ucla.edu}\\
  URL: \url{http://jeroenooms.github.io}
}

\begin{document}


\fvset{listparameters={\setlength{\topsep}{0pt}}}
\renewenvironment{Schunk}{\vspace{\topsep}}{\vspace{\topsep}}


\maketitle

\section{Introduction} 

Modern data collection and analysis pipelines increasingly involve collections
of decoupled components in order to better manage software complexity 
through reusability, modularity, and fault isolation \citep{Wegiel:2010:CTT:1932682.1869479}.
These pipelines are frequently built using different programming 
languages for the different phases of data analysis --- collection,
cleaning, modeling, analysis, post-processing, and
presentation --- in order to take advantage of the unique combination of
performance, speed of development, and library support offered by
different environments and languages.  Each stage of such a data
analysis pipeline may produce intermediate results that need to be
stored in a file, or sent over the network for further processing. 

Given these requirements, how do we safely and efficiently share intermediate results
between different applications, possibly written in different
languages, and possibly running on different computer systems?
In computer programming, \emph{serialization} is the process of
translating data structures, variables, and session state into a
format that can be stored or transmitted and then reconstructed in the
original form later \citep{clinec++}.
Programming
languages such as \proglang{R}, \proglang{Julia}, \proglang{Java}, and \proglang{Python} include built-in
support for serialization, but the default formats 
are usually language-specific and thereby lock the user into a single
environment.  

Data analysts and researchers often use character-separated text formats such
as \texttt{CSV} \citep{shafranovich2005common} to export and import
data. However, anyone who has ever used \texttt{CSV} files will have noticed
that this method has many limitations: it is restricted to tabular data,
lacks type-safety, and has limited precision for numeric values.  Moreover,
ambiguities in the format itself frequently cause problems.  For example,
conventions on which characters is used as separator or decimal point vary by
country.  \emph{Extensible Markup Language} (\texttt{XML}) is another
well-established and widely-supported format with the ability to define just
about any arbitrarily complex schema \citep{nolan2013xml}. However, it pays
for this complexity with comparatively large and verbose messages, and added
complexity at the parsing side (which are somewhat mitigated by the
availability of mature libraries and parsers). Because \texttt{XML} is 
text-based and has no native notion of numeric types or arrays, it usually not a
very practical format to store numeric data sets as they appear in statistical
applications.

A more modern format is \emph{JavaScript ObjectNotation} 
(\texttt{JSON}), which is derived from the object literals of
\proglang{JavaScript}, and already widely-used on the world wide web. 
Several \proglang{R} packages implement functions to parse and generate
\texttt{JSON} data from \proglang{R} objects \citep{rjson,RJSONIO,jsonlite}.
\texttt{JSON} natively supports arrays and four primitive types: numbers, strings,
booleans, and null. However, as it too is a text-based format, numbers are
stored as human-readable decimal notation which is inefficient and
leads to loss of type (double versus integer) and precision. 
A number of binary formats based on \texttt{JSON} have been proposed
that reduce the parsing cost and improve efficiency, but these formats
are not widely supported.  Furthermore, such formats lack a separate
schema for the serialized data and thus still duplicate field names
with each message sent over the network or stored in a file.

Once the data serialization needs of an application become complex
enough, developers typically benefit from the use of an
\emph{interface description language}, or \emph{IDL}.  IDLs like
Protocol Buffers \citep{protobuf}, Apache Thrift, and Apache Avro
provide a compact well-documented schema for cross-language data
structures and efficient binary interchange formats.  Since the schema
is provided separately from the data, the data can be
efficiently encoded to minimize storage costs when
compared with simple ``schema-less'' binary interchange formats.
Many sources compare data serialization formats
and show Protocol Buffers perform favorably to the alternatives; see
\citet{Sumaray:2012:CDS:2184751.2184810} for one such comparison.

This paper describes an \proglang{R} interface to Protocol Buffers,
and is organized as follows. Section~\ref{sec:protobuf}
provides a general high-level overview of Protocol Buffers as well as a basic
motivation for their use.
Section~\ref{sec:rprotobuf-basic} describes the interactive \proglang{R} interface
provided by the \CRANpkg{RProtoBuf} package, and introduces the two main abstractions:
\emph{Messages} and \emph{Descriptors}.  Section~\ref{sec:rprotobuf-classes}
details the implementation details of the main S4 classes and methods.  
Section~\ref{sec:types} describes the challenges of type coercion
between \proglang{R} and other languages.  Section~\ref{sec:evaluation} introduces a
general \proglang{R} language schema for serializing arbitrary \proglang{R} objects and evaluates
it against the serialization capabilities built directly into \proglang{R}.  Sections~\ref{sec:mapreduce}
and \ref{sec:opencpu} provide real-world use cases of \CRANpkg{RProtoBuf}
in MapReduce and web service environments, respectively, before
Section~\ref{sec:summary} concludes.

\section{Protocol Buffers}
\label{sec:protobuf}

Protocol Buffers are a modern, language-neutral, platform-neutral,
extensible mechanism for sharing and storing structured data.  Some of
the key features provided by Protocol Buffers for data analysis include:

\begin{itemize}
\item \emph{Portable}:  Enable users to send and receive data between
  applications as well as different computers or operating systems.
\item \emph{Efficient}:  Data is serialized into a compact binary
  representation for transmission or storage.
\item \emph{Extensible}:  New fields can be added to Protocol Buffer schemas
  in a forward-compatible way that does not break older applications.
\item \emph{Stable}:  Protocol Buffers have been in wide use for over a
  decade.
\end{itemize}

\begin{figure}[bp]
\begin{center}
\includegraphics[width=\textwidth]{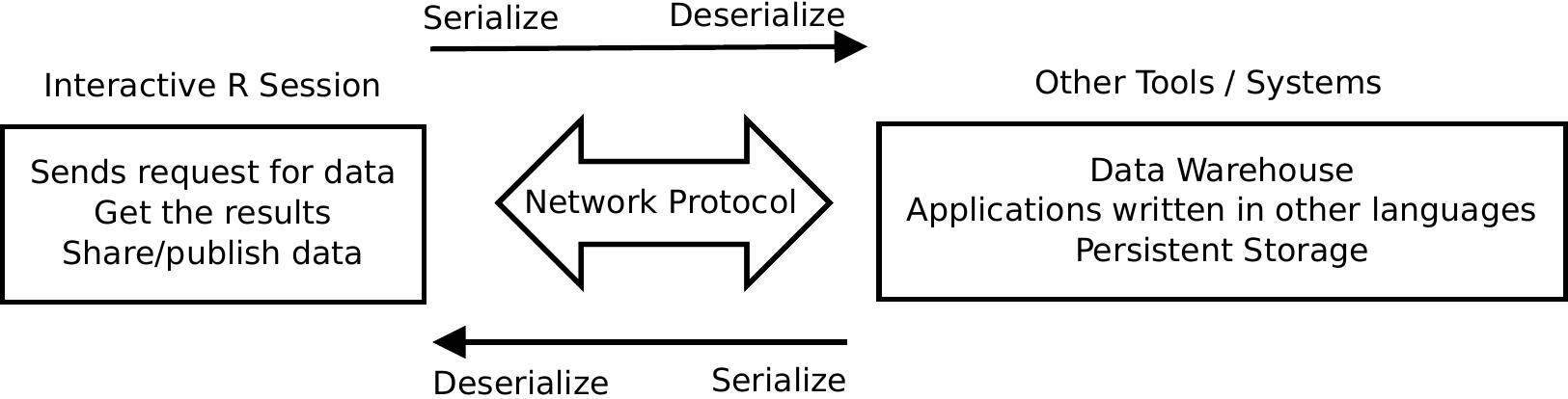}
\end{center}
\caption{Example protobuf usage}
\label{fig:protobuf-distributed-usecase}
\end{figure}

Figure~\ref{fig:protobuf-distributed-usecase} illustrates an example
communication work flow with Protocol Buffers and an interactive \proglang{R} session.
Common use cases include populating a request remote-procedure call (RPC)
Protocol Buffer in \proglang{R} that is then serialized and sent over the network to a
remote server.  The server would then deserialize the message, act on the
request, and respond with a new Protocol Buffer over the network. 
The key difference to, say, a request to an \pkg{Rserve} instance is that
the remote server may be implemented in any language, with no
dependence on \proglang{R}.

While traditional IDLs have at times been criticized for code bloat and
complexity, Protocol Buffers are based on a simple list and records
model that is flexible and easy to use.  The schema for structured
Protocol Buffer data is defined in \texttt{.proto} files, which may
contain one or more message types.  Each message type has one or more
fields.  A field is specified with a unique number (called a \emph{tag number}), a name, a value
type, and a field rule specifying whether the field is optional,
required, or repeated.  The supported value types are numbers,
enumerations, booleans, strings, raw bytes, or other nested message
types.  The \texttt{.proto} file syntax for defining the structure of Protocol
Buffer data is described comprehensively on Google Code\footnote{See 
\url{http://code.google.com/apis/protocolbuffers/docs/proto.html}.}.
Table~\ref{tab:proto} shows an example \texttt{.proto} file that
defines the \texttt{tutorial.Person} type\footnote{The compound name
  \texttt{tutorial.Person} in R is derived from the name of the
  message (\emph{Person}) and the name of the package defined at the top of the
  \texttt{.proto} file in which it is defined (\emph{tutorial}).}.  The \proglang{R} code in the right
column shows an example of creating a new message of this type and
populating its fields.

\noindent
\begin{table}
\begin{tabular}{p{.40\textwidth}p{0.55\textwidth}}
\toprule
Schema : \texttt{addressbook.proto} & Example \proglang{R} Session\\
\cmidrule{1-2}
\begin{minipage}{.40\textwidth}
\vspace{2mm}
\begin{example}
package tutorial;
message Person {
 required string name = 1;
 required int32 id = 2;
 optional string email = 3;
 enum PhoneType {
   MOBILE = 0; HOME = 1;
   WORK = 2;
 }
 message PhoneNumber {
   required string number = 1;
   optional PhoneType type = 2;
 }
 repeated PhoneNumber phone = 4;
}
\end{example}
\vspace{2mm}
\end{minipage} & \begin{minipage}{.55\textwidth}
\begin{Schunk}
\begin{Sinput}
R> library(RProtoBuf)
R> p <- new(tutorial.Person,id=1,name="Dirk")
R> p$name
\end{Sinput}
\begin{Soutput}
[1] "Dirk"
\end{Soutput}
\begin{Sinput}
R> p$name <- "Murray"
R> cat(as.character(p))
\end{Sinput}
\begin{Soutput}
name: "Murray"
id: 1
\end{Soutput}
\begin{Sinput}
R> serialize(p, NULL)
\end{Sinput}
\begin{Soutput}
 [1] 0a 06 4d 75 72 72 61 79 10 01
\end{Soutput}
\begin{Sinput}
R> class(p)
\end{Sinput}
\begin{Soutput}
[1] "Message"
attr(,"package")
[1] "RProtoBuf"
\end{Soutput}
\end{Schunk}
\end{minipage} \\
\bottomrule
\end{tabular}
\caption{The schema representation from a \texttt{.proto} file for the
  \texttt{tutorial.Person} class (left) and simple \proglang{R} code for creating
  an object of this class and accessing its fields (right).}
\label{tab:proto}
\end{table}

For added speed and efficiency, the \proglang{C++}, \proglang{Java},
and \proglang{Python} bindings to
Protocol Buffers are used with a compiler that translates a Protocol
Buffer schema description file (ending in \texttt{.proto}) into
language-specific classes that can be used to create, read, write, and
manipulate Protocol Buffer messages.  The \proglang{R} interface, in contrast,
uses a reflection-based API that makes some operations slightly
slower but which is much more convenient for interactive data analysis.
All messages in \proglang{R} have a single class
structure, but different accessor methods are created at runtime based
on the named fields of the specified message type, as described in the
next section.

\section{Basic Usage: Messages and descriptors}
\label{sec:rprotobuf-basic}

This section describes how to use the \proglang{R} API to create and manipulate
Protocol Buffer messages in \proglang{R}, and how to read and write the
binary representation of the message (often called the \emph{payload}) to files and arbitrary binary
\proglang{R} connections.
The two fundamental building blocks of Protocol Buffers are \emph{Messages}
and \emph{Descriptors}.  Messages provide a common abstract encapsulation of
structured data fields of the type specified in a Message Descriptor.
Message Descriptors are defined in \texttt{.proto} files and define a
schema for a particular named class of messages.

\subsection[Importing message descriptors from .proto files]{Importing message descriptors from \texttt{.proto} files}

To create or parse a Protocol Buffer Message, one must first read in 
the message type specification from a \texttt{.proto} file. The 
\texttt{.proto} files are imported using the \code{readProtoFiles}
function, which can either import a single file, all files in a directory,
or every \texttt{.proto} file provided by a particular \proglang{R} package.

After importing proto files, the corresponding message descriptors are
available by name from the \code{RProtoBuf:DescriptorPool} environment in 
the \proglang{R} search path.  This environment is implemented with the 
user-defined tables framework from the \pkg{RObjectTables} package
available from the OmegaHat project \citep{RObjectTables}.  Instead of
being associated with a static hash table, this environment
dynamically queries the in-memory database of loaded descriptors
during normal variable lookup.

\begin{Schunk}
\begin{Sinput}
R> ls("RProtoBuf:DescriptorPool")
\end{Sinput}
\begin{Soutput}
[1] "rexp.CMPLX"                   "rexp.REXP"                   
[3] "rexp.STRING"                  "rprotobuf.HelloWorldRequest" 
[5] "rprotobuf.HelloWorldResponse" "tutorial.AddressBook"        
[7] "tutorial.Person"             
\end{Soutput}
\end{Schunk}

\subsection{Creating a message}

New messages are created with the \texttt{new} function which accepts
a Message Descriptor and optionally a list of ``name = value'' pairs
to set in the message.

\begin{Schunk}
\begin{Sinput}
R> p1 <- new(tutorial.Person)
R> p <- new(tutorial.Person, name = "Murray", id = 1)
\end{Sinput}
\end{Schunk}

\subsection{Access and modify fields of a message}

Once the message is created, its fields can be queried
and modified using the dollar operator of \proglang{R}, making Protocol
Buffer messages seem like lists.

\begin{Schunk}
\begin{Sinput}
R> p$name
\end{Sinput}
\begin{Soutput}
[1] "Murray"
\end{Soutput}
\begin{Sinput}
R> p$id
\end{Sinput}
\begin{Soutput}
[1] 1
\end{Soutput}
\begin{Sinput}
R> p$email <- "murray@stokely.org"
\end{Sinput}
\end{Schunk}

As opposed to \proglang{R} lists, no partial matching is performed
and the name must be given entirely.
The \verb|[[| operator can also be used to query and set fields
of a messages, supplying either their name or their tag number:

\begin{Schunk}
\begin{Sinput}
R> p[["name"]] <- "Murray Stokely"
R> p[[ 2 ]] <- 3
R> p[["email"]]
\end{Sinput}
\begin{Soutput}
[1] "murray@stokely.org"
\end{Soutput}
\end{Schunk}

Protocol Buffers include a 64-bit integer type, but \proglang{R} lacks native
64-bit integer support.  A workaround is available and described in
Section~\ref{sec:int64} for working with large integer values.

\subsection{Display messages}

Protocol Buffer messages and descriptors implement \texttt{show}
methods that provide basic information about the message:

\begin{Schunk}
\begin{Sinput}
R> p
\end{Sinput}
\begin{Soutput}
[1] "message of type 'tutorial.Person' with 3 fields set"
\end{Soutput}
\end{Schunk}

For additional information, such as for debugging purposes,
the \texttt{as.character} method provides a more complete ASCII
representation of the contents of a message.

\begin{Schunk}
\begin{Sinput}
R> writeLines(as.character(p))
\end{Sinput}
\begin{Soutput}
name: "Murray Stokely"
id: 3
email: "murray@stokely.org"
\end{Soutput}
\end{Schunk}

\subsection{Serializing messages}

One of the primary benefits of Protocol Buffers is the efficient
binary wire-format representation.  
The \texttt{serialize} method is implemented for
Protocol Buffer messages to serialize a message into a sequence of
bytes (raw vector) that represents the message.
The raw bytes can then be parsed back into the original message safely
as long as the message type is known and its descriptor is available.

\begin{Schunk}
\begin{Sinput}
R> serialize(p, NULL)
\end{Sinput}
\begin{Soutput}
 [1] 0a 0e 4d 75 72 72 61 79 20 53 74 6f 6b 65 6c 79 10 03 1a 12 6d 75
[23] 72 72 61 79 40 73 74 6f 6b 65 6c 79 2e 6f 72 67
\end{Soutput}
\end{Schunk}

The same method can be used to serialize messages to files:

\begin{Schunk}
\begin{Sinput}
R> tf1 <- tempfile()
R> serialize(p, tf1)
R> readBin(tf1, raw(0), 500)
\end{Sinput}
\begin{Soutput}
 [1] 0a 0e 4d 75 72 72 61 79 20 53 74 6f 6b 65 6c 79 10 03 1a 12 6d 75
[23] 72 72 61 79 40 73 74 6f 6b 65 6c 79 2e 6f 72 67
\end{Soutput}
\end{Schunk}

Or to arbitrary binary connections:

\begin{Schunk}
\begin{Sinput}
R> tf2 <- tempfile()
R> con <- file(tf2, open = "wb")
R> serialize(p, con)
R> close(con)
R> readBin(tf2, raw(0), 500)
\end{Sinput}
\begin{Soutput}
 [1] 0a 0e 4d 75 72 72 61 79 20 53 74 6f 6b 65 6c 79 10 03 1a 12 6d 75
[23] 72 72 61 79 40 73 74 6f 6b 65 6c 79 2e 6f 72 67
\end{Soutput}
\end{Schunk}

\texttt{serialize} can also be called in a more traditional
object oriented fashion using the dollar operator:

\begin{Schunk}
\begin{Sinput}
R> # serialize to a file
R> p$serialize(tf1)
R> # serialize to a binary connection
R> con <- file(tf2, open = "wb")
R> p$serialize(con)
R> close(con)
\end{Sinput}
\end{Schunk}

\subsection{Parsing messages}

The \CRANpkg{RProtoBuf} package defines the \code{read} and
\code{readASCII} functions to read messages from files, raw vectors,
or arbitrary connections.  \code{read} expects to read the message
payload from binary files or connections and \code{readASCII} parses
the human-readable ASCII output that is created with
\code{as.character}.

The binary representation of the message
does not contain information that can be used to dynamically
infer the message type, so we have to provide this information
to the \code{read} function in the form of a descriptor:

\begin{Schunk}
\begin{Sinput}
R> msg <- read(tutorial.Person, tf1)
R> writeLines(as.character(msg))
\end{Sinput}
\begin{Soutput}
name: "Murray Stokely"
id: 3
email: "murray@stokely.org"
\end{Soutput}
\end{Schunk}

The \texttt{input} argument of \texttt{read} can also be a binary
readable \proglang{R} connection, such as a binary file connection:

\begin{Schunk}
\begin{Sinput}
R> con <- file(tf2, open = "rb")
R> message <- read(tutorial.Person, con)
R> close(con)
R> writeLines(as.character(message))
\end{Sinput}
\begin{Soutput}
name: "Murray Stokely"
id: 3
email: "murray@stokely.org"
\end{Soutput}
\end{Schunk}

Finally, the payload of the message can be used:

\begin{Schunk}
\begin{Sinput}
R> # reading the raw vector payload of the message
R> payload <- readBin(tf1, raw(0), 5000)
R> message <- read(tutorial.Person, payload)
\end{Sinput}
\end{Schunk}

\texttt{read} can also be used as a pseudo-method of the descriptor
object:

\begin{Schunk}
\begin{Sinput}
R> # reading from a file
R> message <- tutorial.Person$read(tf1)
R> # reading from a binary connection
R> con <- file(tf2, open = "rb")
R> message <- tutorial.Person$read(con)
R> close(con)
R> # read from the payload
R> message <- tutorial.Person$read(payload)
\end{Sinput}
\end{Schunk}

\section{Under the hood: S4 classes, methods, and pseudo methods}
\label{sec:rprotobuf-classes}

The \CRANpkg{RProtoBuf} package uses the S4 system to store
information about descriptors and messages.  Using the S4 system
allows the package to dispatch methods that are not
generic in the S3 sense, such as \texttt{new} and
\texttt{serialize}.
Table~\ref{class-summary-table} lists the six
primary Message and Descriptor classes in \CRANpkg{RProtoBuf}.  Each \proglang{R} object
contains an external pointer to an object managed by the
\texttt{protobuf} \proglang{C++} library, and the \proglang{R} objects make calls into more
than 100 \proglang{C++} functions that provide the
glue code between the \proglang{R} language classes and the underlying \proglang{C++}
classes.

\begin{table}[bp]
\centering
\begin{tabular}{lccl}
\toprule
\textbf{Class}      & 
     \textbf{Slots} & 
     \textbf{Methods} & 
     \textbf{Dynamic Dispatch}\\
\cmidrule{1-4}
Message             & 2 & 20 & yes (field names)\\
Descriptor          & 2 & 16 & yes (field names, enum types, nested types)\\
FieldDescriptor     & 4 & 18 & no\\
EnumDescriptor      & 4 & 11 & yes (enum constant names)\\
EnumValueDescriptor & 3 & \phantom{1}6 & no\\
FileDescriptor      & 3 & \phantom{1}6 & yes (message/field definitions)\\
\bottomrule
\end{tabular}
\caption{\label{class-summary-table}Overview of class, slot, method and
  dispatch relationships}
\end{table}

The \CRANpkg{Rcpp} package
\citep{eddelbuettel2011rcpp,eddelbuettel2013seamless} is used to 
facilitate this integration of the \proglang{R} and \proglang{C++} code for these objects.
Each method is wrapped individually which allows us to add 
user-friendly custom error handling, type coercion, and performance
improvements at the cost of a more verbose implementation.
The \CRANpkg{RProtoBuf} package in many ways motivated
the development of \CRANpkg{Rcpp} Modules \citep{eddelbuettel2013exposing},
which provide a more concise way of wrapping \proglang{C++} functions and classes
in a single entity.

The \CRANpkg{RProtoBuf} package supports two forms for calling
functions with these S4 classes:
\begin{itemize}
\item The functional dispatch mechanism of the the form
  \verb|method(object, arguments)| (common to \proglang{R}), and
\item The traditional object oriented notation
  \verb|object$method(arguments)|.
\end{itemize}

Additionally, \CRANpkg{RProtoBuf} supports tab completion for all
classes.  Completion possibilities include pseudo-method names for all
classes, plus \emph{dynamic dispatch} on names or types specific to a given
object.  This functionality is implemented with the
\texttt{.DollarNames} S3 generic function defined in the \pkg{utils}
package.

\subsection{Messages}

The \texttt{Message} S4 class represents Protocol Buffer Messages and
is the core abstraction of \CRANpkg{RProtoBuf}. Each \texttt{Message}
contains a pointer to a \texttt{Descriptor} which defines the schema
of the data defined in the Message, as well as a number of
\texttt{FieldDescriptors} for the individual fields of the message.  A
complete list of the slots and methods for \texttt{Messages}
is available in Table~\ref{Message-methods-table}.

\begin{Schunk}
\begin{Sinput}
R> new(tutorial.Person)
\end{Sinput}
\begin{Soutput}
[1] "message of type 'tutorial.Person' with 0 fields set"
\end{Soutput}
\end{Schunk}

\begin{table}[tbp]
\centering
\begin{small}
\begin{tabular}{lp{10cm}}
\toprule
\textbf{Slot} & \textbf{Description} \\
\cmidrule(r){2-2}
\texttt{pointer} & External pointer to the \texttt{Message} object of the \proglang{C++} protobuf library. Documentation for the
\texttt{Message} class is available from the Protocol Buffer project page. \\
\texttt{type} & Fully qualified name of the message. For example a \texttt{Person} message
has its \texttt{type} slot set to \texttt{tutorial.Person} \\[.3cm]
\textbf{Method} & \textbf{Description} \\
\cmidrule(r){2-2}
\texttt{has} & Indicates if a message has a given field.   \\
\texttt{clone} & Creates a clone of the message \\
\texttt{isInitialized} & Indicates if a message has all its required fields set\\
\texttt{serialize} & serialize a message to a file, binary connection, or raw vector\\
\texttt{clear} & Clear one or several fields of a message, or the entire message\\
\texttt{size} & The number of elements in a message field\\
\texttt{bytesize} & The number of bytes the message would take once serialized\\[3mm]
\texttt{swap} & swap elements of a repeated field of a message\\
\texttt{set} & set elements of a repeated field\\
\texttt{fetch} & fetch elements of a repeated field\\
\texttt{setExtension} & set an extension of a message\\
\texttt{getExtension} & get the value of an extension of a message\\
\texttt{add} & add elements to a repeated field \\[3mm]
\texttt{str} & the \proglang{R} structure of the message\\
\texttt{as.character} & character representation of a message\\
\texttt{toString} & character representation of a message (same as \texttt{as.character}) \\
\texttt{as.list} & converts message to a named \proglang{R} list\\
\texttt{update} & updates several fields of a message at once\\
\texttt{descriptor} & get the descriptor of the message type of this message\\
\texttt{fileDescriptor} & get the file descriptor of this message's descriptor\\
\hline
\end{tabular}
\end{small}
\caption{\label{Message-methods-table}Description of slots and methods for the \texttt{Message} S4 class}
\end{table}

\subsection{Descriptors}

Descriptors describe the type of a Message.  This includes what fields
a message contains and what the types of those fields are.  Message
descriptors are represented in \proglang{R} by the \emph{Descriptor} S4
class. The class contains the slots \texttt{pointer} and
\texttt{type}.  Similarly to messages, the \verb|$| operator can be
used to retrieve descriptors that are contained in the descriptor, or
invoke pseudo-methods.

When \CRANpkg{RProtoBuf} is first loaded it calls
\texttt{readProtoFiles} to read in the example \texttt{addressbook.proto} file
included with the package.  The \texttt{tutorial.Person} descriptor
and all other descriptors defined in the loaded \texttt{.proto} files are
then available on the search path\footnote{This explains why the example in
Table~\ref{tab:proto} lacked an explicit call to
\texttt{readProtoFiles}.}.

\begin{Schunk}
\begin{Sinput}
R> tutorial.Person$email # field descriptor
\end{Sinput}
\begin{Soutput}
[1] "descriptor for field 'email' of type 'tutorial.Person' "
\end{Soutput}
\begin{Sinput}
R> tutorial.Person$PhoneType # enum descriptor
\end{Sinput}
\begin{Soutput}
[1] "descriptor for enum 'PhoneType' of type 'tutorial.Person' with 3 values"
\end{Soutput}
\begin{Sinput}
R> tutorial.Person$PhoneNumber # nested type descriptor
\end{Sinput}
\begin{Soutput}
[1] "descriptor for type 'tutorial.Person.PhoneNumber' "
\end{Soutput}
\begin{Sinput}
R> # same as
R> tutorial.Person.PhoneNumber
\end{Sinput}
\begin{Soutput}
[1] "descriptor for type 'tutorial.Person.PhoneNumber' "
\end{Soutput}
\end{Schunk}

Table~\ref{Descriptor-methods-table} provides a complete list of the
slots and available methods for Descriptors.

\begin{table}[tbp]
\centering
\begin{small}
\begin{tabular}{lp{10cm}}
\toprule
\textbf{Slot} & \textbf{Description} \\
\cmidrule(r){2-2}
\texttt{pointer} & External pointer to the \texttt{Descriptor} object of the \proglang{C++} proto library. Documentation for the
\texttt{Descriptor} class is available from the Protocol Buffer project page.\\
\texttt{type} & Fully qualified path of the message type. \\[.3cm]
\textbf{Method} & \textbf{Description} \\
\cmidrule(r){2-2}
\texttt{new} & Creates a prototype of a message described by this descriptor.\\
\texttt{read} & Reads a message from a file or binary connection.\\
\texttt{readASCII} & Read a message in ASCII format from a file or
text connection.\\
\texttt{name} & Retrieve the name of the message type associated with
this descriptor.\\
\texttt{as.character} & character representation of a descriptor\\
\texttt{toString} & character representation of a descriptor (same as \texttt{as.character}) \\
\texttt{as.list} & return a named
list of the field, enum, and nested descriptors included in this descriptor.\\
\texttt{asMessage} & return DescriptorProto message. \\
\texttt{fileDescriptor} & Retrieve the file descriptor of this
descriptor.\\
\texttt{containing\_type} & Retrieve the descriptor describing the message type containing this descriptor.\\
\texttt{field\_count} & Return the number of fields in this descriptor.\\
\texttt{field} & Return the descriptor for the specified field in this descriptor.\\
\texttt{nested\_type\_count} & The number of nested types in this descriptor.\\
\texttt{nested\_type} & Return the descriptor for the specified nested 
type in this descriptor.\\
\texttt{enum\_type\_count} & The number of enum types in this descriptor.\\
\texttt{enum\_type} & Return the descriptor for the specified enum
type in this descriptor.\\
\bottomrule
\end{tabular}
\end{small}
\caption{\label{Descriptor-methods-table}Description of slots and methods for the \texttt{Descriptor} S4 class}
\end{table}

\subsection{Field descriptors}
\label{subsec-field-descriptor}

The class \emph{FieldDescriptor} represents field
descriptors in \proglang{R}. This is a wrapper S4 class around the
\texttt{google::protobuf::FieldDescriptor} \proglang{C++} class.
Table~\ref{fielddescriptor-methods-table} describes the methods
defined for the \texttt{FieldDescriptor} class.

\begin{table}[tbp]
\centering
\begin{small}
\begin{tabular}{lp{10cm}}
\toprule
\textbf{Slot} & \textbf{Description} \\
\cmidrule(r){2-2}
\texttt{pointer} & External pointer to the \texttt{FieldDescriptor} \proglang{C++} variable \\
\texttt{name} & Simple name of the field \\
\texttt{full\_name} & Fully qualified name of the field \\
\texttt{type} & Name of the message type where the field is declared \\[.3cm]
\textbf{Method} & \textbf{Description} \\
\cmidrule(r){2-2}
\texttt{as.character} & Character representation of a descriptor\\
\texttt{toString} & Character representation of a descriptor (same as \texttt{as.character}) \\
\texttt{asMessage} & Return FieldDescriptorProto message. \\
\texttt{name} & Return the name of the field descriptor.\\
\texttt{fileDescriptor} & Return the fileDescriptor where this field is defined.\\
\texttt{containing\_type} & Return the containing descriptor of this field.\\
\texttt{is\_extension} & Return TRUE if this field is an extension.\\
\texttt{number} & Gets the declared tag number of the field.\\
\texttt{type} & Gets the type of the field.\\
\texttt{cpp\_type} & Gets the \proglang{C++} type of the field.\\
\texttt{label} & Gets the label of a field (optional, required, or repeated).\\
\texttt{is\_repeated} & Return TRUE if this field is repeated.\\
\texttt{is\_required} & Return TRUE if this field is required.\\
\texttt{is\_optional} & Return TRUE if this field is optional.\\
\texttt{has\_default\_value} & Return TRUE if this field has a default value.\\
\texttt{default\_value} & Return the default value.\\
\texttt{message\_type} & Return the message type if this is a message type field.\\
\texttt{enum\_type} & Return the enum type if this is an enum type field.\\
\bottomrule
\end{tabular}
\end{small}
\caption{\label{fielddescriptor-methods-table}Description of slots and
  methods for the \texttt{FieldDescriptor} S4 class}
\end{table}

\subsection{Enum descriptors}
\label{subsec-enum-descriptor}

The class \emph{EnumDescriptor} represents enum descriptors in \proglang{R}.
This is a wrapper S4 class around the
\texttt{google::protobuf::EnumDescriptor} \proglang{C++} class.
Table~\ref{enumdescriptor-methods-table} describes the methods
defined for the \texttt{EnumDescriptor} class.

The \verb|$| operator can be used to retrieve the value of enum
constants contained in the EnumDescriptor, or to invoke
pseudo-methods.

The \texttt{EnumDescriptor} contains information about what values this type
defines, while the \texttt{EnumValueDescriptor} describes an
individual enum constant of a particular type.

\begin{Schunk}
\begin{Sinput}
R> tutorial.Person$PhoneType
\end{Sinput}
\begin{Soutput}
[1] "descriptor for enum 'PhoneType' of type 'tutorial.Person' with 3 values"
\end{Soutput}
\begin{Sinput}
R> tutorial.Person$PhoneType$WORK
\end{Sinput}
\begin{Soutput}
[1] 2
\end{Soutput}
\end{Schunk}

\begin{table}[tbp]
\centering
\begin{small}
\begin{tabular}{lp{10cm}}
\toprule
\textbf{Slot} & \textbf{Description} \\
\cmidrule(r){2-2}
\texttt{pointer} & External pointer to the \texttt{EnumDescriptor} \proglang{C++} variable \\
\texttt{name} & Simple name of the enum \\
\texttt{full\_name} & Fully qualified name of the enum \\
\texttt{type} & Name of the message type where the enum is declared \\[.3cm]
\textbf{Method} & \textbf{Description} \\
\cmidrule(r){2-2}
\texttt{as.list} & return a named
integer vector with the values of the enum and their names.\\
\texttt{as.character} & character representation of a descriptor\\
\texttt{toString} & character
representation of a descriptor (same as \texttt{as.character}) \\
\texttt{asMessage} & return EnumDescriptorProto message. \\
\texttt{name} & Return the name of the enum descriptor.\\
\texttt{fileDescriptor} & Return the fileDescriptor where this field is defined.\\
\texttt{containing\_type} & Return the containing descriptor of this field.\\
\texttt{length} & Return the number of constants in this enum.\\
\texttt{has} & Return TRUE if this enum contains the specified named constant string.\\
\texttt{value\_count} & Return the number of constants in this enum (same as \texttt{length}).\\
\texttt{value} & Return the EnumValueDescriptor of an enum value of specified index, name, or number.\\
\bottomrule
\end{tabular}
\end{small}
\caption{\label{enumdescriptor-methods-table}Description of slots and methods for the \texttt{EnumDescriptor} S4 class}
\end{table}

\subsection{Enum value descriptors}
\label{subsec-enumvalue-descriptor}

The class \emph{EnumValueDescriptor} represents enumeration value
descriptors in \proglang{R}.  This is a wrapper S4 class around the
\texttt{google::protobuf::EnumValueDescriptor} \proglang{C++} class.
Table~\ref{EnumValueDescriptor-methods-table} describes the methods
defined for the \texttt{EnumValueDescriptor} class.

The \verb|$| operator can be used to invoke pseudo-methods.

\begin{Schunk}
\begin{Sinput}
R> tutorial.Person$PhoneType$value(1)
\end{Sinput}
\begin{Soutput}
[1] "enum value descriptor tutorial.Person.MOBILE"
\end{Soutput}
\begin{Sinput}
R> tutorial.Person$PhoneType$value(name="HOME")
\end{Sinput}
\begin{Soutput}
[1] "enum value descriptor tutorial.Person.HOME"
\end{Soutput}
\begin{Sinput}
R> tutorial.Person$PhoneType$value(number=1)
\end{Sinput}
\begin{Soutput}
[1] "enum value descriptor tutorial.Person.HOME"
\end{Soutput}
\end{Schunk}

\begin{table}[tbp]
\centering
\begin{small}
\begin{tabular}{lp{10cm}}
\toprule
\textbf{Slot} & \textbf{Description} \\
\cmidrule(r){2-2}
\texttt{pointer} & External pointer to the \texttt{EnumValueDescriptor} \proglang{C++} variable \\
\texttt{name} & simple name of the enum value \\
\texttt{full\_name} & fully qualified name of the enum value \\[.3cm]
\textbf{Method} & \textbf{Description} \\
\cmidrule(r){2-2}
\texttt{number} & return the number of this EnumValueDescriptor. \\
\texttt{name} & Return the name of the enum value descriptor.\\
\texttt{enum\_type} & return the EnumDescriptor type of this EnumValueDescriptor. \\
\texttt{as.character} & character representation of a descriptor. \\
\texttt{toString} & character representation of a descriptor (same as \texttt{as.character}). \\
\texttt{asMessage} & return EnumValueDescriptorProto message. \\
\bottomrule
\end{tabular}
\end{small}
\caption{\label{EnumValueDescriptor-methods-table}Description of slots
  and methods for the \texttt{EnumValueDescriptor} S4 class}
\end{table}

\subsection{File descriptors}
\label{subsec-file-descriptor}

\begin{table}[tbp]
\centering
\begin{small}
\begin{tabular}{lp{10cm}}
\toprule
\textbf{Slot} & \textbf{Description} \\
\cmidrule(r){2-2}
\texttt{pointer} & external pointer to the \texttt{FileDescriptor} object of the \proglang{C++} proto library. Documentation for the
\texttt{FileDescriptor} class is available from the Protocol Buffer project page:
\url{http://developers.google.com/protocol-buffers/docs/reference/cpp/google.protobuf.descriptor.html#FileDescriptor} \\
\texttt{filename} & fully qualified pathname of the \texttt{.proto} file.\\
\texttt{package} & package name defined in this \texttt{.proto} file.\\[.3cm]
\textbf{Method} & \textbf{Description} \\
\cmidrule(r){2-2}
\texttt{name} & Return the filename for this FileDescriptorProto.\\
\texttt{package} & Return the file-level package name specified in this FileDescriptorProto.\\
\texttt{as.character} & character representation of a descriptor. \\
\texttt{toString} & character representation of a descriptor (same as \texttt{as.character}). \\
\texttt{asMessage} & return FileDescriptorProto message. \\
\texttt{as.list} & return named list of descriptors defined in this file descriptor.\\
\bottomrule
\end{tabular}
\end{small}
\caption{\label{filedescriptor-methods-table}Description of slots and methods for the \texttt{FileDescriptor} S4 class}
\end{table}

The class \emph{FileDescriptor} represents file descriptors in \proglang{R}.
This is a wrapper S4 class around the
\texttt{google::protobuf::FileDescriptor} \proglang{C++} class.
Table~\ref{filedescriptor-methods-table} describes the methods
defined for the \texttt{FileDescriptor} class.

The \verb|$| operator can be used to retrieve named fields defined in
the FileDescriptor, or to invoke pseudo-methods.

\begin{Schunk}
\begin{Sinput}
R> f <- tutorial.Person$fileDescriptor()
R> f
\end{Sinput}
\begin{Soutput}
[1] "file descriptor for package tutorial (/Library/Frameworks/R.framework/Versions/3.0/Resources/library/RProtoBuf/proto/addressbook.proto)"
\end{Soutput}
\begin{Sinput}
R> f$Person
\end{Sinput}
\begin{Soutput}
[1] "descriptor for type 'tutorial.Person' "
\end{Soutput}
\end{Schunk}

\section{Type coercion}
\label{sec:types}

One of the benefits of using an Interface Definition Language (IDL)
like Protocol Buffers is that it provides a highly portable basic type
system. This permits different language and hardware implementations to map to
the most appropriate type in different environments.

Table~\ref{table-get-types} details the correspondence between the
field type and the type of data that is retrieved by \verb|$| and \verb|[[|
extractors.  Three types in particular need further attention due to
specific differences in the \proglang{R} language: booleans, unsigned
integers, and 64-bit integers.

\begin{table}[h]
\centering
\begin{small}
\begin{tabular}{lp{5cm}p{5cm}}
\toprule
Field type & \proglang{R} type (non repeated) & \proglang{R} type (repeated) \\
\cmidrule(r){2-3}
double	& \texttt{double} vector & \texttt{double} vector \\
float	& \texttt{double} vector & \texttt{double} vector \\[3mm]
uint32	  & \texttt{double} vector & \texttt{double} vector \\
fixed32	  & \texttt{double} vector & \texttt{double} vector \\[3mm]
int32	  & \texttt{integer} vector & \texttt{integer} vector \\
sint32	  & \texttt{integer} vector & \texttt{integer} vector \\
sfixed32  & \texttt{integer} vector & \texttt{integer} vector \\[3mm]
int64	  & \texttt{integer} or \texttt{character}
vector    & \texttt{integer} or \texttt{character} vector \\
uint64	  & \texttt{integer} or \texttt{character} vector & \texttt{integer} or \texttt{character} vector \\
sint64	  & \texttt{integer} or \texttt{character} vector & \texttt{integer} or \texttt{character} vector \\
fixed64	  & \texttt{integer} or \texttt{character} vector & \texttt{integer} or \texttt{character} vector \\
sfixed64  & \texttt{integer} or \texttt{character} vector & \texttt{integer} or \texttt{character} vector \\[3mm]
bool	& \texttt{logical} vector & \texttt{logical} vector \\[3mm]
string	& \texttt{character} vector & \texttt{character} vector \\
bytes	& \texttt{character} vector & \texttt{character} vector \\[3mm]
enum & \texttt{integer} vector & \texttt{integer} vector \\[3mm]
message & \texttt{S4} object of class \texttt{Message} & \texttt{list} of \texttt{S4} objects of class \texttt{Message} \\
\bottomrule
\end{tabular}
\end{small}
\caption{\label{table-get-types}Correspondence between field type and
  \proglang{R} type retrieved by the extractors. Note that \proglang{R} lacks native
  64-bit integers, so the \code{RProtoBuf.int64AsString} option is
  available to return large integers as characters to avoid losing
  precision.  This option is described in Section~\ref{sec:int64}.}
\end{table}

\subsection{Booleans}

\proglang{R} booleans can accept three values: \texttt{TRUE}, \texttt{FALSE}, and
\texttt{NA}.  However, most other languages, including the Protocol
Buffer schema, only accept \texttt{TRUE} or \texttt{FALSE}.  This means
that we simply can not store \proglang{R} logical vectors that include all three
possible values as booleans.  The library will refuse to store
\texttt{NA}s in Protocol Buffer boolean fields, and users must instead
choose another type (such as enum or integer) capable of storing three
distinct values.

\begin{CodeChunk}
\begin{CodeInput}
R> a <- new(protobuf_unittest.TestAllTypes)
R> a$optional_bool <- TRUE
R> a$optional_bool <- FALSE
R> a$optional_bool <- NA
\end{CodeInput}
\begin{CodeOutput}
Error: NA boolean values can not be stored in bool Protocol Buffer fields
\end{CodeOutput}
\end{CodeChunk}

\subsection{Unsigned integers}

\proglang{R} lacks a native unsigned integer type.  Values between $2^{31}$ and
$2^{32} - 1$ read from unsigned integer Protocol Buffer fields must be
stored as doubles in \proglang{R}.

\begin{Schunk}
\begin{Sinput}
R> as.integer(2^31-1)
\end{Sinput}
\begin{Soutput}
[1] 2147483647
\end{Soutput}
\begin{Sinput}
R> as.integer(2^31 - 1) + as.integer(1)
\end{Sinput}
\begin{Soutput}
[1] NA
\end{Soutput}
\begin{Sinput}
R> 2^31
\end{Sinput}
\begin{Soutput}
[1] 2.147e+09
\end{Soutput}
\begin{Sinput}
R> class(2^31)
\end{Sinput}
\begin{Soutput}
[1] "numeric"
\end{Soutput}
\end{Schunk}

\subsection{64-bit integers}
\label{sec:int64}

\proglang{R} also does not support the native 64-bit integer type. Numeric vectors
with values $\geq 2^{31}$ can only be stored as doubles, which have
limited precision. Thereby \proglang{R} loses the ability to distinguish some
distinct integers:

\begin{Schunk}
\begin{Sinput}
R> 2^53 == (2^53 + 1)
\end{Sinput}
\begin{Soutput}
[1] TRUE
\end{Soutput}
\end{Schunk}

However, most modern languages do have support for 64-bit integers, 
which becomes problematic when \CRANpkg{RProtoBuf} is used to exchange data 
with a system that requires this integer type. To work around this, 
\CRANpkg{RProtoBuf} allows users to get and set 64-bit integer values by specifying 
them as character strings.

If we try to set an int64 field in \proglang{R} to double values, we lose
precision:

\begin{Schunk}
\begin{Sinput}
R> test <- new(protobuf_unittest.TestAllTypes)
R> test$repeated_int64 <- c(2^53, 2^53+1)
R> length(unique(test$repeated_int64))
\end{Sinput}
\begin{Soutput}
[1] 1
\end{Soutput}
\end{Schunk}

But when the values are specified as character strings, \CRANpkg{RProtoBuf}
will automatically coerce them into a true 64-bit integer types 
before storing them in the Protocol Buffer message:

\begin{Schunk}
\begin{Sinput}
R> test$repeated_int64 <- c("9007199254740992", "9007199254740993")
\end{Sinput}
\end{Schunk}

When reading the value back into \proglang{R}, numeric types are returned by
default, but when the full precision is required a character value
will be returned if the \code{RProtoBuf.int64AsString} option is set
to \texttt{TRUE}.  The character values are useful because they can
accurately be used as unique identifiers and can easily be passed to \proglang{R}
packages such as \CRANpkg{int64} \citep{int64} or \CRANpkg{bit64}
\citep{bit64} which represent 64-bit integers in \proglang{R}.

\begin{Schunk}
\begin{Sinput}
R> options("RProtoBuf.int64AsString" = FALSE)
R> test$repeated_int64
\end{Sinput}
\begin{Soutput}
[1] 9.007e+15 9.007e+15
\end{Soutput}
\begin{Sinput}
R> length(unique(test$repeated_int64))
\end{Sinput}
\begin{Soutput}
[1] 1
\end{Soutput}
\begin{Sinput}
R> options("RProtoBuf.int64AsString" = TRUE)
R> test$repeated_int64
\end{Sinput}
\begin{Soutput}
[1] "9007199254740992" "9007199254740993"
\end{Soutput}
\begin{Sinput}
R> length(unique(test$repeated_int64))
\end{Sinput}
\begin{Soutput}
[1] 2
\end{Soutput}
\end{Schunk}

\section[Converting R data structures into Protocol Buffers]{Converting \proglang{R} data structures into Protocol Buffers}
\label{sec:evaluation}

The previous sections discussed functionality in the \CRANpkg{RProtoBuf} package
for creating, manipulating, parsing, and serializing Protocol Buffer
messages of a defined schema.  This is useful when there are
pre-existing systems with defined schemas or significant software
components written in other languages that need to be accessed from
within \proglang{R}.
The package also provides methods for converting arbitrary \proglang{R} data structures into Protocol
Buffers and vice versa with a universal \proglang{R} object schema. The \code{serialize\_pb} and \code{unserialize\_pb}
functions serialize arbitrary \proglang{R} objects into a universal Protocol Buffer 
message:

\begin{Schunk}
\begin{Sinput}
R> msg <- serialize_pb(iris, NULL)
R> identical(iris, unserialize_pb(msg))
\end{Sinput}
\begin{Soutput}
[1] TRUE
\end{Soutput}
\end{Schunk}

In order to accomplish this, \CRANpkg{RProtoBuf} uses the same catch-all \texttt{proto}
schema used by \pkg{RHIPE} for exchanging \proglang{R} data with Hadoop \citep{rhipe}. This 
schema, which we will refer to as \texttt{rexp.proto}, is printed in
the appendix.
The Protocol Buffer messages generated by \CRANpkg{RProtoBuf} and
\pkg{RHIPE} are naturally compatible between the two systems because they use the 
same schema. This shows the power of using a schema-based cross-platform format such
as Protocol Buffers: interoperability is achieved without effort or close coordination.

The \texttt{rexp.proto} schema supports all main \proglang{R} storage types holding \emph{data}.
These include \texttt{NULL}, \texttt{list} and vectors of type \texttt{logical}, 
\texttt{character}, \texttt{double}, \texttt{integer}, and \texttt{complex}. In addition,
every type can contain a named set of attributes, as is the case in \proglang{R}. The \texttt{rexp.proto}
schema does not support some of the special \proglang{R} specific storage types, such as \texttt{function},
\texttt{language} or \texttt{environment}. Such objects have no native equivalent 
type in Protocol Buffers, and have little meaning outside the context of \proglang{R}.
When serializing \proglang{R} objects using \texttt{serialize\_pb}, values or attributes of
unsupported types are skipped with a warning. If the user really wishes to serialize these 
objects, they need to be converted into a supported type. For example, the  can use 
\texttt{deparse} to convert functions or language objects into strings, or \texttt{as.list}
for environments.

\subsection[Evaluation: Converting R data sets]{Evaluation: Converting \proglang{R} data sets}

To illustrate how this method works, we attempt to convert all of the built-in 
data sets from \proglang{R} into this serialized Protocol Buffer representation.

\begin{Schunk}
\begin{Sinput}
R> datasets <- as.data.frame(data(package="datasets")$results)
R> datasets$name <- sub("\\s+.*$", "", datasets$Item)
R> n <- nrow(datasets)
\end{Sinput}
\end{Schunk}

There are 206 standard data sets included in the \pkg{datasets}
package included with \proglang{R}. These data sets include data frames, matrices, time series, tables lists,
and some more exotic data classes. The \texttt{can\_serialize\_pb} method is 
used to determine which of those can fully be converted to the \texttt{rexp.proto}
Protocol Buffer representation. This method simply checks if any of the values or
attributes in an object is of an unsupported type:

\begin{Schunk}
\begin{Sinput}
R> m <- sum(sapply(datasets$name, function(x) can_serialize_pb(get(x))))
\end{Sinput}
\end{Schunk}

192 data sets can be converted to Protocol Buffers
without loss of information (93\%). Upon closer
inspection, all other data sets are objects of class \texttt{nfnGroupedData}.
This class represents a special type of data frame that has some additional 
attributes (such as a \emph{formula} object) used by the \pkg{nlme} package \citep{nlme}.
Because formulas are \proglang{R} \emph{language} objects, they have little meaning to
other systems, and are not supported by the \texttt{rexp.proto} descriptor.
When \texttt{serialize\_pb} is used on objects of this class, it will serialize
the data frame and all attributes, except for the formula.

\begin{Schunk}
\begin{Sinput}
R> attr(CO2, "formula")
\end{Sinput}
\begin{Soutput}
uptake ~ conc | Plant
<environment: R_EmptyEnv>
\end{Soutput}
\begin{Sinput}
R> msg <- serialize_pb(CO2, NULL)
R> object <- unserialize_pb(msg)
R> identical(CO2, object)
\end{Sinput}
\begin{Soutput}
[1] FALSE
\end{Soutput}
\begin{Sinput}
R> identical(class(CO2), class(object))
\end{Sinput}
\begin{Soutput}
[1] TRUE
\end{Soutput}
\begin{Sinput}
R> identical(dim(CO2), dim(object))
\end{Sinput}
\begin{Soutput}
[1] TRUE
\end{Soutput}
\begin{Sinput}
R> attr(object, "formula")
\end{Sinput}
\begin{Soutput}
list()
NULL
\end{Soutput}
\end{Schunk}

\subsection{Compression performance}
\label{sec:compression}

This section compares how many bytes are used to store data sets
using four different methods:

\begin{itemize}
\item normal \proglang{R} serialization \citep{serialization},
\item \proglang{R} serialization followed by gzip,
\item normal Protocol Buffer serialization, and
\item Protocol Buffer serialization followed by gzip.
\end{itemize}

Table~\ref{tab:compression} shows the sizes of 50 sample \proglang{R} data sets as
returned by object.size() compared to the serialized sizes.
Note that Protocol Buffer serialization results in slightly
smaller byte streams compared to native \proglang{R} serialization in most cases,
but this difference disappears if the results are compressed with gzip.
One takeaway from this table is that the universal \proglang{R} object schema
included in \CRANpkg{RProtoBuf} does not in general provide
any significant saving in file size compared to the normal serialization
mechanism in \proglang{R}.
The benefits of \CRANpkg{RProtoBuf} accrue more naturally in applications where
multiple programming languages are involved, or when a more concise
application-specific schema has been defined.  The example in the next
section satisfies both of these conditions.

\begin{table}[h!]
\begin{center}
  \small
\scalebox{0.9}{
\begin{tabular}{lrrrrr}
  \toprule
  Data Set & object.size & \multicolumn{2}{c}{\proglang{R} Serialization} &
  \multicolumn{2}{c}{RProtoBuf Serial.} \\
  & & default & gzipped & default & gzipped \\
  \cmidrule(r){2-6}
  uspop & 584 & 268 & 172 & 211 & 148 \\
  Titanic & 1960 & 633 & 257 & 481 & 249 \\
  volcano & 42656 & 42517 & 5226 & 42476 & 4232 \\
  euro.cross & 2728 & 1319 & 910 & 1207 & 891 \\
  attenu & 14568 & 8234 & 2165 & 7771 & 2336 \\
  ToothGrowth & 2568 & 1486 & 349 & 1239 & 391 \\
  lynx & 1344 & 1028 & 429 & 971 & 404 \\
  nottem & 2352 & 2036 & 627 & 1979 & 641 \\
  sleep & 2752 & 746 & 282 & 483 & 260 \\
  co2 & 4176 & 3860 & 1473 & 3803 & 1453 \\
  austres & 1144 & 828 & 439 & 771 & 410 \\
  ability.cov & 1944 & 716 & 357 & 589 & 341 \\
  EuStockMarkets & 60664 & 59785 & 21232 & 59674 & 19882 \\
  treering & 64272 & 63956 & 17647 & 63900 & 17758 \\
  freeny.x & 1944 & 1445 & 1311 & 1372 & 1289 \\
  Puromycin & 2088 & 813 & 306 & 620 & 320 \\
  warpbreaks & 2768 & 1231 & 310 & 811 & 343 \\
  BOD & 1088 & 334 & 182 & 226 & 168 \\
  sunspots & 22992 & 22676 & 6482 & 22620 & 6742 \\
  beaver2 & 4184 & 3423 & 751 & 3468 & 840 \\
  anscombe & 2424 & 991 & 375 & 884 & 352 \\
  esoph & 5624 & 3111 & 548 & 2240 & 665 \\
  PlantGrowth & 1680 & 646 & 303 & 459 & 314 \\
  infert & 15848 & 14328 & 1172 & 13197 & 1404 \\
  BJsales & 1632 & 1316 & 496 & 1259 & 465 \\
  stackloss & 1688 & 917 & 293 & 844 & 283 \\
  crimtab & 7936 & 4641 & 713 & 1655 & 576 \\
  LifeCycleSavings & 6048 & 3014 & 1420 & 2825 & 1407 \\
  Harman74.cor & 9144 & 6056 & 2045 & 5861 & 2070 \\
  nhtemp & 912 & 596 & 240 & 539 & 223 \\
  faithful & 5136 & 4543 & 1339 & 4936 & 1776 \\
  freeny & 5296 & 2465 & 1518 & 2271 & 1507 \\
  discoveries & 1232 & 916 & 199 & 859 & 180 \\
  state.x77 & 7168 & 4251 & 1754 & 4068 & 1756 \\
  pressure & 1096 & 498 & 277 & 427 & 273 \\
  fdeaths & 1008 & 692 & 291 & 635 & 272 \\
  euro & 976 & 264 & 186 & 202 & 161 \\
  LakeHuron & 1216 & 900 & 420 & 843 & 404 \\
  mtcars & 6736 & 3798 & 1204 & 3633 & 1206 \\
  precip & 4992 & 1793 & 813 & 1615 & 815 \\
  state.area & 440 & 422 & 246 & 405 & 235 \\
  attitude & 3024 & 1990 & 544 & 1920 & 561 \\
  randu & 10496 & 9794 & 8859 & 10441 & 9558 \\
  state.name & 3088 & 844 & 408 & 724 & 415 \\
  airquality & 5496 & 4551 & 1241 & 2874 & 1294 \\
  airmiles & 624 & 308 & 170 & 251 & 148 \\
  quakes & 33112 & 32246 & 9898 & 29063 & 11595 \\
  islands & 3496 & 1232 & 563 & 1098 & 561 \\
  OrchardSprays & 3600 & 2164 & 445 & 1897 & 483 \\
  WWWusage & 1232 & 916 & 274 & 859 & 251 \\
  \bottomrule
  Relative Size & 100\% & 83.7\% & 25.3\% & 80.1\% & 25.6\%\\
  \bottomrule
\end{tabular}
}
\caption{Serialization sizes for default serialization in \proglang{R} and
  \CRANpkg{RProtoBuf} for 50 \proglang{R} data sets.}
\label{tab:compression}
\end{center}
\end{table}

\section{Application: Distributed data collection with MapReduce}
\label{sec:mapreduce}

Many large data sets in fields such as particle physics and information
processing are stored in binned or histogram form in order to reduce
the data storage requirements \citep{scott2009multivariate}.  In the
last decade, the MapReduce programming model \citep{dean2008mapreduce}
has emerged as a popular design pattern that enables the processing of
very large data sets on large compute clusters.

Many types of data analysis over large data sets may involve very rare
phenomenon or deal with highly skewed data sets or inflexible
raw data storage systems from which unbiased sampling is not feasible.
In such situations, MapReduce and binning may be combined as a
pre-processing step for a wide range of statistical and scientific
analyses \citep{blocker2013}.

There are two common patterns for generating histograms of large data
sets in a single pass with MapReduce.  In the first method, each
mapper task generates a histogram over a subset of the data that it
has been assigned, serializes this histogram and sends it to one or
more reducer tasks which merge the intermediate histograms from the
mappers.

\begin{figure}[h!]
\begin{center}
\includegraphics[width=\textwidth]{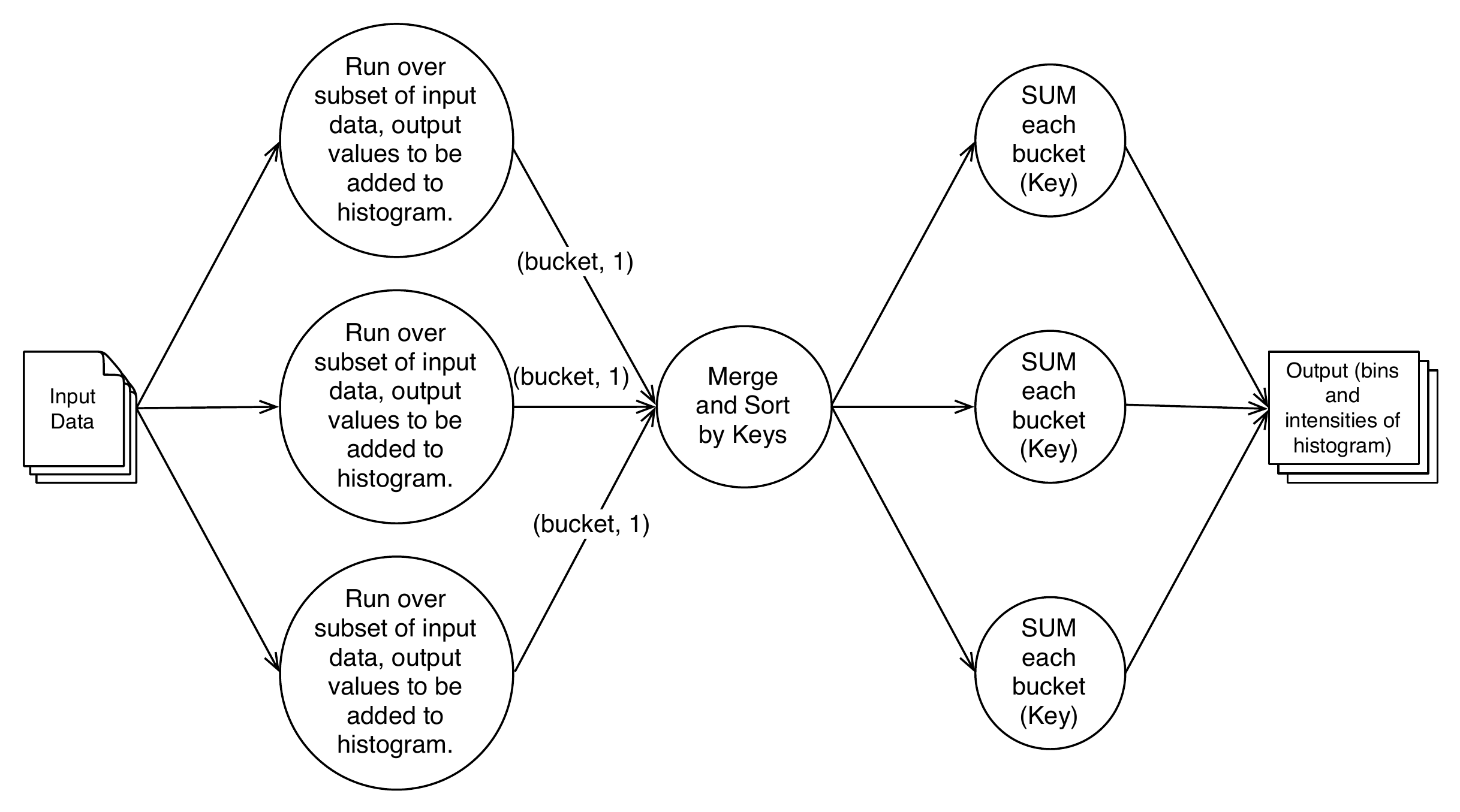}
\end{center}
\caption{Diagram of MapReduce histogram generation pattern}
\label{fig:mr-histogram-pattern1}
\end{figure}

In the second method, illustrated in
Figure~\ref{fig:mr-histogram-pattern1}, each mapper rounds a data
point to a bucket width and outputs that bucket as a key and '1' as a
value.  Reducers then sum up all of the values with the same key and
output to a data store.

In both methods, the mapper tasks must choose identical bucket
boundaries in advance if we are to construct the histogram in a single
pass, even though they are analyzing disjoint parts of the input set
that may cover different ranges.  All distributed tasks involved in
the pre-processing as well as any downstream data analysis tasks must
share a schema of the histogram representation to coordinate
effectively.

The \CRANpkg{HistogramTools} package \citep{histogramtools} enhances
\CRANpkg{RProtoBuf} by providing a concise schema for \proglang{R} histogram objects:

\begin{example}
package HistogramTools;

message HistogramState {
  repeated double breaks = 1;
  repeated int32 counts = 2;
  optional string name = 3;
}
\end{example}

This HistogramState message type is designed to be helpful if some of
the Map or Reduce tasks are written in \proglang{R}, or if those components are
written in other languages and only the resulting output histograms
need to be manipulated in \proglang{R}.  For example, to create HistogramState
messages in Python for later consumption by \proglang{R}, we first compile the 
\texttt{histogram.proto} descriptor into a python module using the
\texttt{protoc} compiler:

\begin{verbatim}
  protoc histogram.proto --python_out=.
\end{verbatim}
This generates a Python module called \texttt{histogram\_pb2.py}, containing both the 
descriptor information as well as methods to read and manipulate the histogram 
message data.  The following simple Python script uses this generated
module to create a histogram and write out the Protocol Buffer
representation to a file:

\begin{Code}
from histogram_pb2 import HistogramState;

# Create empty Histogram message
hist = HistogramState()

# Add breakpoints and binned data set.
hist.counts.extend([2, 6, 2, 4, 6])
hist.breaks.extend(range(6))
hist.name="Example Histogram Created in Python"

# Output the histogram
outfile = open("/tmp/hist.pb", "wb")
outfile.write(hist.SerializeToString())
outfile.close()
\end{Code}

The Protocol Buffer can then be read into \proglang{R} and converted to a native
\proglang{R} histogram object for plotting:

\begin{Code}
library(RProtoBuf)
library(HistogramTools)

# Read the Histogram schema
readProtoFiles(package="HistogramTools")

# Read the serialized histogram file.
hist <- HistogramTools.HistogramState$read("/tmp/hist.pb")
hist
[1] "message of type 'HistogramTools.HistogramState' with 3 fields set"

# Convert to native R histogram object and plot
plot(as.histogram(hist))
\end{Code}

\begin{center}
\includegraphics{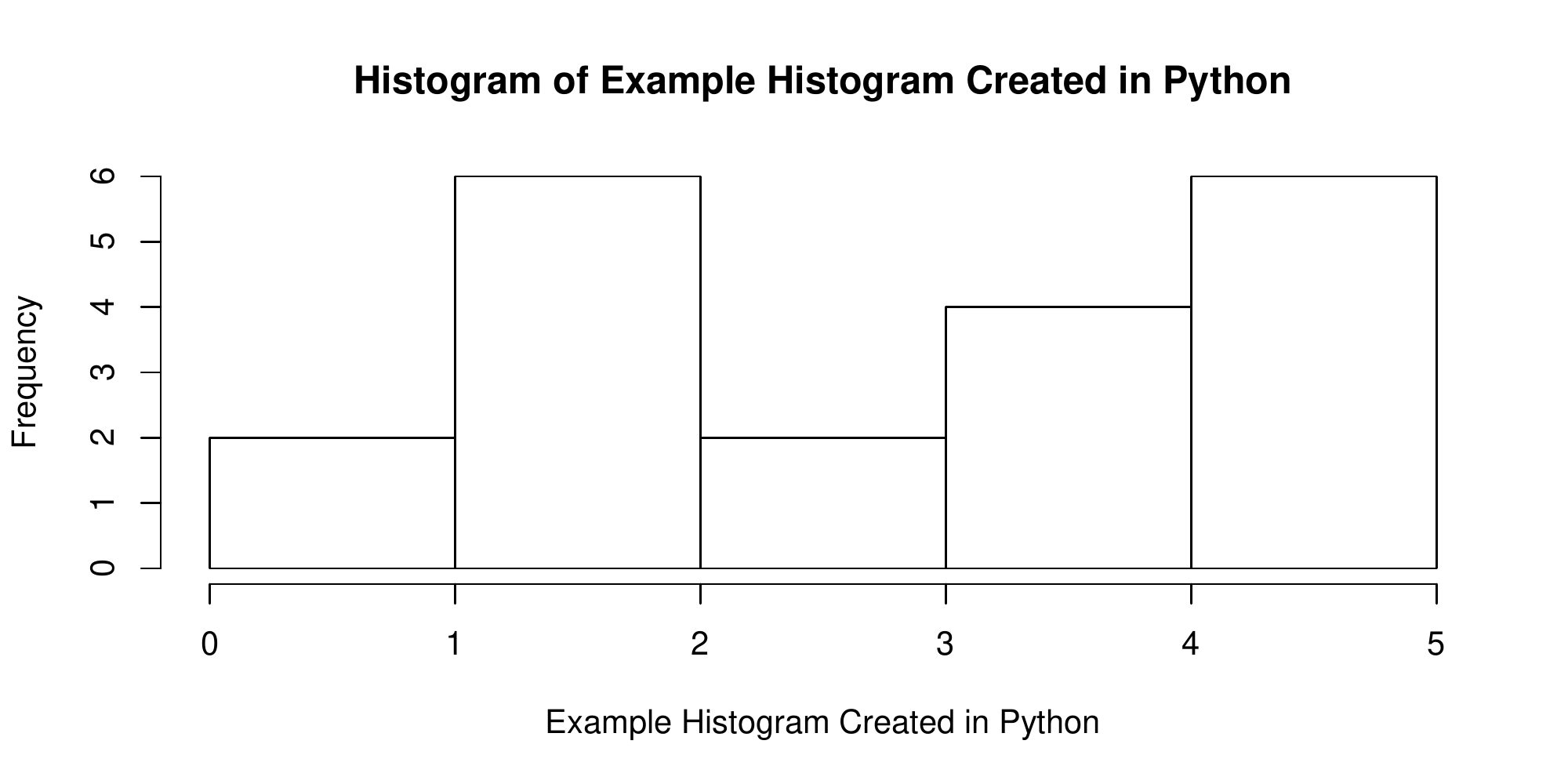}
\end{center}

One of the authors has used this design pattern for several
large-scale studies of distributed storage systems
\citep{sciencecloud,janus}.

\section{Application: Data Interchange in web Services}
\label{sec:opencpu}

As described earlier, the primary application of Protocol Buffers is data
interchange in the context of inter-system communications.  Network protocols
such as HTTP provide mechanisms for client-server communication, i.e., how to
initiate requests, authenticate, send messages, etc.  However, network
protocols generally do not regulate the \emph{content} of messages: they
allow transfer of any media type, such as web pages, static files or
multimedia content.  When designing systems where various components require
exchange of specific data structures, we need something on top of the network
protocol that prescribes how these structures are to be represented in
messages (buffers) on the network. Protocol Buffers solve exactly this
problem by providing a cross-platform method for serializing arbitrary
structures into well defined messages, which can then be exchanged using any
protocol. The descriptors (\texttt{.proto} files) are used to formally define
the interface of a remote API or network application. Libraries to parse and
generate protobuf messages are available for many programming languages,
making it relatively straightforward to implement clients and servers.

\subsection[Interacting with R through HTTPS and Protocol Buffers]{Interacting with \proglang{R} through HTTPS and Protocol Buffers}

One example of a system that supports Protocol Buffers to interact
with \proglang{R} is OpenCPU \citep{opencpu}. OpenCPU is a framework for embedded statistical 
computation and reproducible research based on \proglang{R} and \LaTeX. It exposes a 
HTTP(S) API to access and manipulate \proglang{R} objects and allows for performing 
remote \proglang{R} function calls. Clients do not need to understand 
or generate any \proglang{R} code: HTTP requests are automatically mapped to 
function calls, and arguments/return values can be posted/retrieved
using several data interchange formats, such as Protocol Buffers.  
OpenCPU uses the \texttt{serialize\_pb} and \texttt{unserialize\_pb} functions
from the \CRANpkg{RProtoBuf} package to convert between \proglang{R} objects and protobuf
messages. Therefore, clients need the \texttt{rexp.proto} descriptor mentioned
earlier to parse and generate protobuf messages when interacting with OpenCPU.

\subsection[HTTP GET: Retrieving an R object]{HTTP GET: Retrieving an \proglang{R} object}

The \texttt{HTTP GET} method is used to read a resource from OpenCPU. For example,
to access the data set \texttt{Animals} from the package \texttt{MASS}, a 
client performs the following HTTP request:

\begin{verbatim}
  GET https://public.opencpu.org/ocpu/library/MASS/data/Animals/pb
\end{verbatim}
The postfix \texttt{/pb} in the URL tells the server to send this
object in the form of a protobuf message. Alternative formats include 
\texttt{/json}, \texttt{/csv}, \texttt{/rds} and others. If the request
is successful, OpenCPU returns the serialized object with HTTP status 
code 200 and HTTP response header \texttt{Content-Type: application/x-protobuf}. 
The latter is the conventional MIME type that formally notifies the client to
interpret the response as a protobuf message. 

Because both HTTP and Protocol Buffers have libraries available for many 
languages, clients can be implemented in just a few lines of code. Below
is example code for both \proglang{R} and Python that retrieves a data set from \proglang{R} with 
OpenCPU using a protobuf message. In \proglang{R}, we use the HTTP client from 
the \texttt{httr} package \citep{httr}. In this example we
download a data set which is part of the base \proglang{R} distribution, so we can
verify that the object was transferred without loss of information.

\begin{Schunk}
\begin{Sinput}
R> # Load packages
R> library(RProtoBuf)
R> library(httr)
R> # Retrieve and parse message
R> req <- GET('https://public.opencpu.org/ocpu/library/MASS/data/Animals/pb')
R> output <- unserialize_pb(req$content)
R> # Check that no information was lost
R> identical(output, MASS::Animals)
\end{Sinput}
\end{Schunk}

This code suggests a method for exchanging objects between \proglang{R} servers, however this might as 
well be done without Protocol Buffers. The main advantage of using an inter-operable format 
is that we can actually access \proglang{R} objects from within another
programming language. For example, in a very similar fashion we can retrieve the same
data set in a Python client. To parse messages in Python, we first compile the 
\texttt{rexp.proto} descriptor into a python module using the \texttt{protoc} compiler:

\begin{verbatim}
  protoc rexp.proto --python_out=.
\end{verbatim}
This generates Python module called \texttt{rexp\_pb2.py}, containing both the 
descriptor information as well as methods to read and manipulate the \proglang{R} object 
message. In the example below we use the HTTP client from the \texttt{urllib2}
module. 

\begin{verbatim}
# Import modules
import urllib2
from rexp_pb2 import REXP

# Retrieve message
req = urllib2.Request('https://public.opencpu.org/ocpu/library/MASS/data/Animals/pb')
res = urllib2.urlopen(req)
        
# Parse rexp.proto message
msg = REXP()
msg.ParseFromString(res.read())
print(msg)
\end{verbatim}
The \texttt{msg} object contains all data from the Animals data set. From here we
can easily extract the desired fields for further use in Python.

\subsection[HTTP POST: Calling an R function]{HTTP POST: Calling an \proglang{R} function}

The example above shows how the \texttt{HTTP GET} method retrieves a 
resource from OpenCPU, for example an \proglang{R} object. The \texttt{HTTP POST} 
method on the other hand is used for calling functions and running scripts, 
which is the primary purpose of the framework. As before, the \texttt{/pb} 
postfix requests to retrieve the output as a protobuf message, in this
case the function return value. However, OpenCPU allows us to supply the
arguments of the function call in the form of protobuf messages as well.
This is a bit more work, because clients needs to both generate messages 
containing \proglang{R} objects to post to the server, as well as retrieve and parse
protobuf messages returned by the server. Using Protocol Buffers to post
function arguments is not required, and for simple (scalar) arguments 
the standard \texttt{application/x-www-form-urlencoded} format might be sufficient.
However, with Protocol Buffers the client can perform function calls with
more complex arguments such as \proglang{R} vectors or lists. The result is a complete
RPC system to do arbitrary \proglang{R} function calls from within 
any programming language.

The following example \proglang{R} client code performs the remote function call 
\texttt{stats::rnorm(n=42, mean=100)}. The function arguments (in this
case \texttt{n} and \texttt{mean}) as well as the return value (a vector
with 42 random numbers) are transferred using a protobuf message. RPC in
OpenCPU works like the \texttt{do.call} function in \proglang{R}, hence all arguments
are contained within a list.

\begin{Schunk}
\begin{Sinput}
R> #requires httr >= 0.2.99
R> library(httr)
R> library(RProtoBuf)
R> args <- list(n=42, mean=100)
R> payload <- serialize_pb(args, NULL)
R> req <- POST (
+    url = "https://public.opencpu.org/ocpu/library/stats/R/rnorm/pb",
+    body = payload,
+    add_headers (
+      "Content-Type" = "application/x-protobuf"
+    )
+  )
R> #This is the output of stats::rnorm(n=42, mean=100)
R> output <- unserialize_pb(req$content)
R> print(output)
\end{Sinput}
\end{Schunk}
The OpenCPU server basically performs the following steps to process the above RPC request:  

\begin{Schunk}
\begin{Sinput}
R> fnargs <- unserialize_pb(inputmsg)
R> val <- do.call(stats::rnorm, fnargs)
R> outputmsg <- serialize_pb(val)
\end{Sinput}
\end{Schunk}

\section{Summary}  
\label{sec:summary}
Over the past decade, many formats for interoperable
data exchange have become available, each with their unique features,
strengths and weaknesses.  
Text based formats such as \texttt{CSV} and \texttt{JSON} are easy to use, and will likely 
remain popular among statisticians for many years to come. However, in the 
context of increasingly complex analysis stacks and applications involving 
distributed computing as well as mixed language analysis pipelines, choosing a more 
sophisticated data interchange format may reap considerable benefits. 
The Protocol Buffers standard and library offer a unique combination of features, 
performance, and maturity, that seems particularly well suited for data-driven 
applications and numerical computing.

The \CRANpkg{RProtoBuf} package builds on the Protocol Buffers \proglang{C++} library, 
and extends the \proglang{R} system with the ability to create, read,
write, parse, and manipulate Protocol
Buffer messages. \CRANpkg{RProtoBuf} has been used extensively inside Google 
for the past three years by statisticians, analysts, and software engineers.
At the time of this writing there are over 300 active
users of \CRANpkg{RProtoBuf} using it to read data from and otherwise interact
with distributed systems written in \proglang{C++}, \proglang{Java}, \proglang{Python}, and 
other languages. We hope that making Protocol Buffers available to the
\proglang{R} community will contribute towards better software integration
and allow for building even more advanced applications and analysis pipelines 
with \proglang{R}.

\section*{Acknowledgments}

The first versions of \CRANpkg{RProtoBuf} were written during 2009-2010.
Very significant contributions, both in code and design, were made by
Romain Fran\c{c}ois whose continued influence on design and code is
greatly appreciated. Several features of the package reflect
the design of the \CRANpkg{rJava} package by Simon Urbanek.
The user-defined table mechanism, implemented by Duncan Temple Lang for the
purpose of the \pkg{RObjectTables} package, allows for the dynamic symbol lookup.
Kenton Varda was generous with his time in reviewing code and explaining
obscure Protocol Buffer semantics.  Karl Millar was very
helpful in reviewing code and offering suggestions.  Saptarshi Guha's
work on RHIPE and implementation of a universal message type for \proglang{R}
language objects allowed us to add the \code{serialize_pb} and \code{unserialize_pb}
methods for turning arbitrary R objects into Protocol Buffers without
a specialized pre-defined schema.

\newpage
\appendix
\setcounter{secnumdepth}{0}

\section*{Appendix: The rexp.proto schema descriptor}
\label{rexp.proto}

Below a print of the \texttt{rexp.proto} schema (originally designed by \cite{rhipe})
that is included with the \CRANpkg{RProtoBuf} package and used by \texttt{serialize\_pb} and
\texttt{unserialize\_pb}.

\begin{verbatim}
package rexp;

message REXP {
  enum RClass {
    STRING = 0;
    RAW = 1;
    REAL = 2;
    COMPLEX = 3;
    INTEGER = 4;
    LIST = 5;
    LOGICAL = 6;
    NULLTYPE = 7;
  }
  enum RBOOLEAN {
    F=0;
    T=1;
    NA=2;
  }

  required RClass rclass = 1 ; 
  repeated double realValue = 2 [packed=true];
  repeated sint32 intValue = 3 [packed=true];
  repeated RBOOLEAN booleanValue = 4;
  repeated STRING stringValue = 5;
  optional bytes rawValue = 6;
  repeated CMPLX complexValue = 7;
  repeated REXP rexpValue = 8;
  repeated string attrName = 11;
  repeated REXP attrValue = 12;
}
message STRING {
  optional string strval = 1;
  optional bool isNA = 2 [default=false];
}
message CMPLX {
  optional double real = 1 [default=0];
  required double imag = 2;
}
\end{verbatim}
\newpage
\bibliography{article}

\end{document}